\documentclass[aip,jmp,preprint,showpacs]{revtex4-1}
\usepackage[utf8x]{inputenc}
\usepackage[T1]{fontenc}
\usepackage[english]{babel}
\usepackage{eurosym}
\usepackage{amsmath, amsthm, amscd, amssymb,amsbsy,amsfonts,wasysym}
\usepackage{lastpage}
\usepackage[dvips]{graphicx}
\usepackage{fancyhdr}
\usepackage{vmargin}
\usepackage{multirow,framed}
\usepackage[table]{xcolor}
\usepackage{makeidx}
\usepackage{color}

\begin{document}

\title{Effective non-Markovian description of a system interacting with a bath}   
\author{L. Ferialdi}
\email{ferialdi@math.lmu.de}
\affiliation{Mathematisches Institut, Ludwig-Maximilians-Universit\"at, Theresienstr. 39, 80333 Munich.}
\author{D. D\"urr}
\email{duerr@math.lmu.de}
\affiliation{Mathematisches Institut, Ludwig-Maximilians-Universit\"at, Theresienstr. 39, 80333 Munich.}
\date{\today}
\begin{abstract}
We study a harmonic system coupled to a chain of first neighbor interacting oscillators. After deriving the exact dynamics of the system, we prove that one can effectively describe the exact dynamics by considering a suitable shorter chain. We provide the explicit expression for such an effective dynamics and we provide an upper bound on the error one makes considering it instead of the dynamics of the full chain. We eventually prove how error, timescale and number of modes in the truncated chain are related.
\end{abstract}
\pacs{03.65.Yz, 42.50.Lc, 03.65.Ta}
\maketitle

\section{Introduction}


Every quantum system unavoidably interacts with the surrounding environment. The dynamics describing such open quantum systems are in general very difficult to treat as  they involve the whole history of the process. In order to analyze the physical features of these systems one needs to consider effective descriptions, which have a simpler structure and are then more suitable for a detailed analysis. If the timescale of the system is much smaller than the one of the environment, one can use the Markovian approximation. Markovian dynamics are the most widely used, and their mathematical and physical features very well known~\cite{BrePet02}. However, many physical systems in chemistry and biology, do not display such a clear separation of timescales and they cannot be described by Markovian dynamics~\cite{Pometal,Xiaetal,Thoetal}. In order to study such non-Markovian systems, one then needs to look for different kind of effective descriptions, not based one a timescale selection.
The main problem one has to face when dealing with open systems is that the environment has a very large number of degrees of freedom which are hard to control. A possible solution to this problem is to consider an effective description which involves a small number of degrees of freedom. In this paper we prove that one can effectively describe the influence of a big environment, with an effective environment made of a (much) smaller number of constituents.

The independent oscillators (IO) model, is one of the most widely used to describe open quantum systems~\cite{Zwa73,CalLeg83,ForLewOco88a}. In this model, the environment is described by a set of independent harmonic oscillators which are linearly coupled to the relevant system. The IO model has been widely studied, and it allowed for the description of Markovian and non-Markovian quantum Brownian motion~\cite{CalLeg83,GraTal83,HuPazZha92}. Furthermore, one can derive a Generalized Langevin Equation, which is suitable for a phenomenological description of thermal and diffusive properties~\cite{Mor65,Kub66,BenKac81,ForKac87,ForLewOco88b}. However, though the GLE shows the existence of non-Markovian effects, it does not allow for an effective description of these.
In a recent paper~\cite{FerDur15}, it has been shown that a chain model for the environment suits better for this goal. In this model, which is widely used also in other fields~\cite{Rub60,ForKacMaz65}, the environment is described by a chain of first neighbor interacting oscillators. The ordering in the interaction (absent in the IO model), gives a clearer physical picture of how the non-Markovian effects build up and allows for an effective description of the dynamics.

A first attempt of effectively describing the short time dynamics of a system has been made in~\cite{GinBurCed06}. Defining some \lq\lq collective modes\rq\rq, the authors obtain an effective Hamiltonian and they show that the cumulant expansion of this is equivalent to that of the original Hamiltonian up to the third term. Though they can argue the equivalence of the two dynamics for \lq\lq short times\rq\rq, this result is not rigorously proven, and it does not give the error made by considering the effective dynamics.

A step forward the derivation of an effective dynamics has been taken in~\cite{FerDur15}, where the authors solved the Heisenberg equations of motion for the chain models obtaining a description equivalent to the GLE. Unlike the GLE, which cannot be analytically solved for any environment, the dynamics obtained in~\cite{FerDur15} is completely general and its structure paves the way for an effective description. The aim of this paper is to complete this program: we prove that, for a harmonic system, one can derive an effective dynamics for any set of chain parameters, giving an explicit bound on the error one makes considering the effective dynamics instead of the full one.

The paper is organized as follows: in Section II we summarize the main result of~\cite{FerDur15}, giving a mathematical account of it; in Section III we prove that one can give an effective description of the full dynamics, we prove that the error one makes is bounded. We eventually provide the relationship among timescale, number of modes and error.

\section{Non-Markovian dynamics of an harmonic oscillator}
Our starting model is a system interacting linearly with an environment of $N$ independent oscillators (IO). Such a model is described by the following Hamiltonian:
\begin{equation}\label{Hio}
H_{\text{\tiny{IO}}}=\frac{p^2}{2M}+V(x)+x\sum_{k=1}^N c_k q_k+\sum_{k=1}^N \frac{1}{2}\left(p_k^2+\omega_k^2q_k^2\right)
\end{equation}
where at this stage $V(x)$ is a generic  potential, while later we will focus on harmonic systems, i.e. purely quadratic potentials, which one may see, if one wishes to do so, as Taylor expansion of the generic potential. Since we are considering only short times such a Taylor expansion might be justifiable, but we do not do that in this paper.  $x,p$ are the position and momentum operators of the relevant system, and $q_k, p_k$ are position and momentum operators of the environmental oscillators. Such oscillators have proper frequency $\omega_k$ and they are coupled to the system via the positive constants $c_k$.  

The Heisenberg equations of motion for the environmental oscillators are easily written as follows:
\begin{equation}\label{motio}
\frac{d^2}{dt^2}{\bf q}(t)=-\boldsymbol{\omega}\cdot{\bf q}(t)
\end{equation}
where ${\mathbf q^{T}}=(q_1,\ldots q_N)$ is the vector of the environmental position operators, and $\boldsymbol{\omega}$ is the diagonal matrix of the oscillators frequencies: $\boldsymbol{\omega}=\mathrm{diag}(\omega_1^2,\ldots\omega_N^2)$, with $\omega_1<\omega_2<\dots\omega_N$.

As we have already mentioned, a chain model for the bath is more suitable to study the short time behavior of the dynamics. Such a model is described by the following Hamiltonian:
\begin{equation}\label{Hchain}
H_{\text{\tiny{CHAIN}}}=\frac{p^2}{2M}+V(x)+DxX_1+\sum_{k=2}^N D_{k-1} X_{k-1}X_k+\sum_{k=1}^N \frac{1}{2}\left(P_k^2+\Omega_k^2X_k^2\right)
\end{equation}
The chain oscillators have position $X_k$, momentum $P_k$, proper frequency $\Omega_k$, and display a first-neighbor interaction with positive coupling $D_k$. 
The Heisenberg equations for $X_k$ read
\begin{equation}\label{motchain}
\frac{d^2}{dt^2}{\bf X}(t)=-\boldsymbol{T}\cdot{\bf X}(t)
\end{equation}
where ${\mathbf X^{T}}=(X_1,\ldots X_N)$, and ${\mathbf T}$ is the following tridiagonal matrix:
\begin{equation}\label{tri}
{\mathbf T}=\left(
\begin{array}{cccc}
\Omega_1^2& -D_1& 0&\dots\\
-D_1&\Omega_2^2 &-D_2&\dots\\
0&-D_2& \Omega_3^2&\dots\\
\vdots &\vdots &\vdots & \ddots
\end{array}\right)
\end{equation}
In order for the chain description to be equivalent to the initial IO model, one needs to require that the dynamics given by~\eqref{Hio} and~\eqref{Hchain} are equivalent. Accordingly, the parameters entering~\eqref{Hchain} are not free, but particular combinations of the parameters of~\eqref{Hio}, and the chain oscillators $X$ must be specific linear combination of the independent $q$. We define
\begin{equation}\label{X}
X_j= \sum_k O_{jk}q_k\,.
\end{equation}
where $\mathbf{O}$ is an orthogonal $N\times N$ matrix.  
The condition of equivalent dynamics is fulfilled only when the systems of equations of motions~\eqref{motio} and~\eqref{motchain} are equivalent. Substituting~\eqref{X}, one finds that this is true 
when
\begin{equation}\label{IEP}
\mathbf{T}=\mathbf{O}\cdot\boldsymbol{\omega}\mathbf{O}^{T}
\end{equation}
The problem of determining a matrix starting from its eigenvalues is known with the name of Inverse Eigenvalue Problem~\cite{Gla04,ChuGol05}. For our case of study, such problem can be exactly solved, i.e. one can analytically determine the entries of $\mathbf{O}$, and through them the matrix $\mathbf{T}$ and the parameters of the chain $\Omega_k, D_k$. For the details of the solution of the IEP and the derivation of the parameters one can refer to~\cite{FerDur15}, here we summarize the result.
The entries of the matrix $\mathbf O$ read 
\begin{equation}\label{O}
O_{jk}=\left(\prod_{l=1}^{j-1}D^{-1}_{j-1}\right)P_{j-1}(\omega_k)\,,
\end{equation}
where $P_j(\lambda)$ is the characteristic polynomial of the $j$-th leading principal minor of $\mathbf T$, evaluated in $\lambda$. The explicit expressions for the $P_j$ are determined recursively exploiting the following recurrence relation:
\begin{equation}\label{recP}
P_{j+1}(\lambda)= (\Omega_j^2-\lambda)P_j(\lambda)-D_j^2P_{j-1}(\lambda)
\end{equation}
with $P_{-1}=0$.
Once the transformation matrix ${\mathbf O}$ is determined, the entries of ${\mathbf T}$ are given by the following relations:
\begin{eqnarray}
\Omega_j^2&=&\sum_k\omega_k^2O_{jk}^2\\
D_j&=&-\sum_k\omega_k^2O_{jk}O_{j+1k}\,.
\end{eqnarray}
From now on we will consider the matrix ${\mathbf T}$ as known, i.e. as fully determined in terms of the parameters of the IO model, and we will assume the dynamics given by Eq.~\eqref{Hchain} to be equivalent to that given by Eq.~\eqref{Hio}. 

\subsection{Non-Markovian dynamics of an harmonic oscillator}

After having established the equivalence between the two models, we determine how the chain modes affect the dynamics of the relevant system. At this purpose, we need to solve the set of equations of motion~\eqref{motchain}. Since the equations of motion are equivalent for a quantum or classical systems, we treat the latter case in order to keep the treatment easier. We consider the relevant system to be an harmonic oscillator with proper frequency $\Omega$, in such a way that~\eqref{motchain} can be explicitly rewritten as follows:
\begin{eqnarray}\label{mot}
\frac{d^2}{dt^2}x(t)&=&-\Omega^2x(t)+D X_1(t)\\
\frac{d^2}{dt^2}X_{i}(t)&=&-\Omega_i^2X_{i}(t)+D_{i-1}X_{j-1}(t)+D_iX_{i+1}(t)\,,\quad 1\leq i\leq N-1\\
\frac{d^2}{dt^2}X_{N}(t)&=&-\Omega_N^2X_{N}(t)+D_{N-1}X_{N-1}(t)\,,
\end{eqnarray}
We independently solve the equations of the previous system, and rewrite them as follows: 
\begin{eqnarray}\label{intmot}
x(t)&=&f_0+\int_0^t \frac{\sin[\Omega_0(t-s)]}{\Omega_0} DX_{1}(s)ds\\
\label{intmoti}X_i(t)&=& f_i(t)\!+\!\!\int_0^t \!\frac{\sin[\Omega_i(t-s)]}{\Omega_i}\! (D_{i-1} X_{i-1}(s)\!+\!D_i X_{i+1}(s))ds,\,1\leq i\leq N\!-\!1\\
\label{intmotN}X_{N}(t)&=&f_N+\int_0^t \frac{\sin[\Omega_N(t-s)]}{\Omega_N} D_{N-1} X_{N-1}(s)ds\,,
\end{eqnarray}
where  we have relabeled $\Omega_0=\Omega$, $X_0=x$, and
\begin{equation}\label{init}
 f_i(t)=X_i(0)\cos[\Omega_it]+\dot{X}_i(0)\frac{\sin[\Omega_i t]}{\Omega_i}\,.
\end{equation}

In order to obtain an equation for $x$ in terms of the $X_i$, we substitute recursively in $x(t)$ the equations for $X_1(t),X_2(t),\ldots X_n(t)$, and we prove the following Theorem.

\vspace{0.5cm}
\noindent{\bf Theorem 1:} 
Let $\{x(t),X_i(t)\}_{i=1,\dots,N}\in\mathbb{R}$ be the set of functions solving the system~\eqref{intmot}-\eqref{intmotN}. Let the functions $K_i(t-s):\mathbb{R}^2\rightarrow\mathbb{R}$ and $\tilde{f}_i(t):\mathbb{R}^2\rightarrow\mathbb{R}$ be defined recursively as follows $\forall 1\leq i\leq N$:
\begin{eqnarray}
\label{Ki} K_{i}(t-s)&=&\int_s^tK_{i-1}(t-l)\sin[\Omega_i(l-s)]dl\\
\label{tilf} \tilde{f}_i(t)&=&\tilde{f}_{i-1}(t)+\left(\prod_{l=0}^{i-1}\frac{D_l}{\Omega_l}\right)\int_0^t K_{i-1}(t-s)f_{i}(s)ds
\end{eqnarray}
with $K_0(t-l)=\sin[\Omega(t-l)]$, $\tilde{f}_0(t)=f_0(t)$, and $f_i(s)$ given by Eq.~\eqref{init}. Hence, $\forall n\leq N$, Eq.~\eqref{intmot} for $x(t)$ can be rewritten as follows:
\begin{eqnarray}\label{xn}
x(t)&=&\tilde{f}_n(t)+\sum_{i=1}^n\left(\prod_{l=0}^i\frac{D_l}{\Omega_l}\right)\frac{D_{i-1}}{D_i}\int_0^t K_{i}(t-s)X_{i-1}(s)ds\nonumber\\
&&+\left(\prod_{l=0}^n\frac{D_l}{\Omega_l}\right)\int_0^t K_{n}(t-s)X_{n+1}(s)ds\,.
\end{eqnarray}

\noindent{\bf Proof:} The proof is by induction. First of all we show that for $n=1$, Eq.~\eqref{xn} is correct. Substituting Eq.~\eqref{intmoti} for $X_1(t)$ in Eq.~\eqref{intmot} one finds
\begin{eqnarray}
x(t)&=&f_{0}(t)+\frac{D}{\Omega}\int_0^t \sin[\Omega(t-s)]f_{1}(s)ds\nonumber\\
&&+\frac{D^2}{\Omega\Omega_1}\int_0^t K_{1}(t-s)x(s)ds+\frac{DD_1}{\Omega\Omega_1}\int_0^t K_{1}(t-s)X_{2}(s)ds\,.
\end{eqnarray}
This equation can be easily recast in a form like Eq.~\eqref{xn}. Assume now that Eq.~\eqref{xn} is true for a generic $n\leq N$, and substitute Eq.~\eqref{intmoti} for $X_{n+1}(s)$ in the second line of Eq.~\eqref{xn}. After some simple manipulation one can show that the Eq.~\eqref{xn} reads now as follows:
\begin{eqnarray}\label{xn+1}
x(t)&=&\tilde{f}_{n+1}(t)+\sum_{i=1}^{n+1}\left(\prod_{l=0}^i\frac{D_l}{\Omega_l}\right)\frac{D_{i-1}}{D_i}\int_0^t K_{i}(t-s)X_{i-1}(s)ds\nonumber\\
&&+\left(\prod_{l=0}^{n+1}\frac{D_l}{\Omega_l}\right)\int_0^t K_{n+1}(t-s)X_{n+2}(s)ds
\end{eqnarray}
that is exactly Eq.~\eqref{xn} for $n\rightarrow n+1$. This completes the proof. $\square$

\vspace{0.2cm}
It is important to note that equation~\eqref{xn} is exact. Although only the equations for the first $n$ $X_i$ have been substituted into $x(t)$, the term depending on $X_{n+1}$ of Eq.~\eqref{xn} encodes  the information regarding the evolution of the remaining $N-n$ modes. It is easy to prove that, if one substitutes the equations for all the $N$ $X_i$, the following Corollary holds true:

\vspace{0.5cm}
\noindent{\bf Corollary:} Let the hypothesis of Theorem 1 be true, and let $n=N$. Hence,
\begin{equation}\label{xN}
x(t)=\tilde{f}_N(t)+\sum_{i=1}^N\left(\prod_{l=0}^i\frac{D_l}{\Omega_l}\right)\frac{D_{i-1}}{D_i}\int_0^t K_{i}(t-s)X_{i-1}(s)ds
\end{equation}

\noindent{\bf Proof:} By definition, $D_N=0$. Accordingly, the second line of Eq.~\eqref{xn} is null. $\square$

\vspace{0.2cm}
This equation determines how the dynamics of $x(t)$ is affected by the full set of $X_i$: $\tilde{f}_{N}(t)$ accounts for the initial conditions of the collective modes, while the second term is a purely non-Markovian contribution which involves the whole past evolution of the collective modes.
It is important to underline that each $X_i$ contributes to $x(t)$ via $K_{i+1}$. This is a crucial feature of the dynamics that will play a fundamental role in the following calculations.

In Eqs.~\eqref{xn} and \eqref{xN} the dependence on $x(t)$ is not completely explicit. Recalling that $X_0=x$ we rewrite such equations as follows:
\begin{equation}\label{xNint}
x(t)=\frac{D^2}{\Omega\Omega_1}\int_0^t K_{1}(t-s)x(s)ds+F_N(t)
\end{equation}
where the function $F_N(t)$ collecting all terms that do not depend on $x$ is defined as follows:
\begin{eqnarray}
\label{FN1} F_N(t)&=&\tilde{f}_n(t)+\sum_{i=2}^N\left(\prod_{l=0}^i\frac{D_l}{\Omega_l}\right)\frac{D_{i-1}}{D_i}\int_0^t K_{i}(t-s)X_{i-1}(s)ds\nonumber\\
&&+\left(\prod_{l=0}^n\frac{D_l}{\Omega_l}\right)\int_0^t K_{n}(t-s)X_{n+1}(s)ds\\
\label{FN2} &=&\tilde{f}_N(t)+\sum_{i=2}^N\left(\prod_{l=0}^i\frac{D_l}{\Omega_l}\right)\frac{D_{i-1}}{D_i}\int_0^t K_{i}(t-s)X_{i-1}(s)ds\,,
\end{eqnarray}
where the definitions come from Eq.~\eqref{xn} and Eq.~\eqref{xN} respectively. Note that if one chooses $n=N$ in the first equation, one recovers the second one as expected.
Eq.~\eqref{xNint} explicitly shows that the dynamics of $x$ is ruled by an integral equation. Since the kernel $K_1(t-s)$ is a linear combination of two sine functions, Eq.~\eqref{xNint} can be solved using standard techniques~\cite{PolMan08}. The solution reads
\begin{equation}\label{xsol}
x(t)=F_N(t)+\frac{D^2}{\mu_1\mu_2(\mu_2^2-\mu_1^2)}\int_0^t \left(\mu_2\sin[\mu_1(t-s)]-\mu_1\sin[\mu_2(t-s)]\right)F_N(s)ds
\end{equation}
where
\begin{equation}
\mu_{1,2}=\sqrt{\frac{1}{2}\left(\Omega^2+\Omega_1^2\pm\sqrt{\Delta}\right)}\,,\qquad\Delta=(\Omega^2-\Omega_1^2)^2-4D^2
\end{equation}
In order to avoid multi-valued $x(t)$, $\mu_{1,2}$ have to be real, i.e the condition $\Delta\geq 0$ has to hold true.
~Equation~\eqref{xsol} is the exact solution of our problem, as it displays the dynamics of an harmonic oscillator under the influence of a chain of $N$ harmonic oscillators. One can identify two different type of contributions: a diffusive one, given by those terms contained in $F_N(t)$ that depend on the initial conditions $q_k(0)$; and a purely non-Markovian one, given by the integral terms, and which depends on the interaction among the chain modes. This equation gives a description of non-Markovian dynamics which is equivalent to the Generalized Langevin Equation~\cite{FerDur15}. Furthermore, Eq.~\eqref{xsol}  plays an important role because it has the added value of allowing for an effective description of non-Markovian dynamics, i.e. a description in terms of a smaller number of degrees of freedom.

\section{Effective non-Markovian dynamics}
In the study of open quantum system the number $N$ of constituents of the environment is typically assumed to be very large. Controlling the dynamics of so many constituents is not possible. One then has to consider approximated dynamics, or in many practical applications one needs to exploit numerical methods. 
Aim of this section is to understand whether it is possible to describe with a good approximation the full dynamics of an open system, by a particle interacting with a smaller bath (in our case, a shorter chain). In fact, we will prove that there always exist a time-scale such that the true dynamics $x(t)$ of the system is well approximated by truncating the original chain after $n\leq N$ oscillators. 
Let $x_{\text{\tiny{(n)}}}(t)$ denote the the exact evolution of $x$ interacting with a chain of $n$ oscillators. Observe that truncating the chain after the $n$-th oscillator corresponds to setting $D_n=0$. 
Accordingly, the evolution of $x_{\text{\tiny{(n)}}}(t)$ is given by Eq.~\eqref{xsol} with $N$ replaced by $n$. It proves useful for the forthcoming discussion to define the following function:
\begin{equation}\label{eps1}
\epsilon_1(n,t):=F_N(t)-F_n(t)
\end{equation}
Comparing Eq.~\eqref{FN1} for $F_N(t)$, and Eq.~\eqref{FN2} for $F_n(t)$, one can easily show that
\begin{equation}
\mathcal{\epsilon}_{1}(n,t)=\left(\prod_{l=0}^n\frac{D_l}{\Omega_l}\right)\int_0^t K_{n}(t-s)X_{n+1}(s)ds
\end{equation}

We define a further function that measures the error made by considering the truncated dynamics instead of the full one:
\begin{equation}
\epsilon(n,t):=\left|x(t)-x_{\text{\tiny{(n)}}}(t)\right|
\end{equation}
Exploiting Eq.~\eqref{xsol} one finds that
\begin{equation}\label{err}
\epsilon(n,t)=\left|\epsilon_1(n,t)+\epsilon_2(n,t)\right|
\end{equation}
where
\begin{equation}\label{eps2}
\epsilon_2(n,t)=\frac{D^2}{\mu_1\mu_2(\mu_2^2-\mu_1^2)}\int_0^t \left(\mu_2\sin[\mu_1(t-s)]-\mu_1\sin[\mu_2(t-s)]\right)\epsilon_1(n,s)ds
\end{equation}
Our aim is to obtain an upper bound on the error function $\epsilon(n,t)$. Since $\epsilon(n,t)$ strongly depends on the features of the kernels $K_n(t-s)$, in order to reach our goal we need to understand their features.

\subsection{Kernels structure}
 From the definition of Eq.~\eqref{Ki}, one easily understands that each $K_i$ consists of $i$ nested integrals of sine functions. This structure, that is due to the harmonic feature of the chain, turns out to be crucial for the analysis of the system. Indeed, the $k$-th derivatives of $K_i$ have the following remarkable properties.
 
\vspace{0.5cm}
\noindent{\bf Theorem 2:} Let $K_i(t-s):\mathbb{R}^2\rightarrow\mathbb{R}$ be the kernels defined in Eq.~\eqref{Ki}, and let $K_i^{(k)}(t-s)$ be their $k$-th derivatives with respect to $t$. Hence, the following equations hold true:
\begin{eqnarray}
\label{Kik} K_i^{(k)}(t-s)&=&\int_s^t K_{i-1}^{(k)}(t-l)\sin[\Omega_i(l-s)]dl\qquad\forall k\leq2i-1\\
\label{Kikf} K_i^{(2k)}(t-s)&=&\sin[\Omega_i(t-s)]\sum_{j=1}^{k}(-1)^{j+k}K^{(2j-1)}_{i-1}(0)\Omega_i^{2k-2j}\nonumber\\
&&+\int_s^t K_{i-1}^{(2i+1)}(t-l)\sin[\Omega_i(l-s)]dl\qquad\forall k\geq i\\
\label{Kiko} K_i^{(2k-1)}(t-s)&=&\cos[\Omega_i(t-s)]\sum_{j=1}^{k-1}(-1)^{j+k-1}K^{(2j-1)}_{i-1}(0)\Omega_i^{2k-1-2j}\nonumber\\
&&+\int_s^t K_{i-1}^{(2i+1)}(t-l)\sin[\Omega_i(l-s)]dl\qquad\forall k\geq i+1
\end{eqnarray}

\noindent{\bf Proof:} The proof is by induction. We recall the definition of $K_1(t-s)$:
\begin{equation}
K_{1}(t-s)=\int_s^tK_{0}(t-l)\sin[\Omega_1(l-s)]dl=\int_s^t\sin[\Omega(t-l)]\sin[\Omega_1(l-s)]dl
\end{equation}
Differentiating this equation it is easy to check that $K_1^{(1)}$, $K_1^{(2)}$ and $K_1^{(3)}$ satisfy the system above. The proof for higher derivatives can easily be done differentiating recursively.
Assume now that the system~\eqref{Kik}-\eqref{Kiko} holds true, and note that this implies that $K_{i}^{(k)}(0)=0$ for all $k\leq 2i-1$. Accordingly, if one iteratively differentiates the definition~\eqref{Ki} for $K_{i+1}$, one finds
\begin{equation}
K_{i+1}^{(k)}(t-s)=\int_s^t K_{i}^{(k)}(t-l)\sin[\Omega_i(l-s)]dl\qquad\forall k\leq 2i+1
\end{equation}
that is Eq.~\eqref{Kik} for $K_{i+1}$. Differentiating this equation one can easily prove that Eqs.~\eqref{Kikf}-\eqref{Kiko} hold true for $K_{i+1}$ as well. $\square$

\vspace{0.2cm}
This Theorem shows some interesting features of the kernels derivatives which will prove essential. First of all, $K_{i}^{(k)}(0)=0$ not only for all $k\leq 2i-1$, but also for all even $k$. Moreover, iterating Eq.~\eqref{Kiko} one can show that
\begin{equation}\label{k0}
K_i^{(2k-1)}(0)=\left(\prod_{l=0}^i\Omega_l\right)\sum_{\substack{\alpha_j=0\\\sum \alpha_j=k-1-i}}^{k-1-i}\left(\prod_{l=0}^j\Omega_l^{2\alpha_l}\right)\,.
\end{equation}

We exploit Theorem 2  to understand the time behavior of the kernels, by expanding $K_i$ in Taylor series. First of all note that, since the first $2i$ and all the even derivatives of $K_i$ are null, the kernels Taylor series start with the index $i-1$, and they display only odd terms.
\begin{equation}\label{taykern}
K_i(t-s)=\sum_{k=0}^{\infty}\frac{K_i^{(k)}(0)}{k!}(t-s)^k=\sum_{k=i+1}^{\infty}\frac{K_i^{(2k-1)}(0)}{(2k-1)!}(t-s)^{2k-1}\,,
\end{equation}
 This equation gives an interesting insight on the evolution of $x(t)$ and on how the non-Markovian behavior emerges. We recall Eq.~\eqref{xN} which shows that each $X_i$ influences $x(t)$ via $K_{i+1}$. Accordingly, if we consider $t<1$, the smaller $i$ the earlier $X_i$ contributes to the dynamics of $x$. At the beginning, only $X_1$ gives a relevant contribution to the non-Markovian term of the dynamics of $x(t)$, i.e. the integral term of Eq.~\eqref{xN}, and  as time grows, also other modes enter into the game. In other words, for short time scales only the first oscillators of the chain contribute to the non-Markovian dynamics. 

\subsection{Short-time approximation}
From the discussion above one can also infer another fundamental property of the system: if the time scale is short enough, further oscillators can be neglected since their contribution is very small. Accordingly, one can effectively describe the dynamics of the full system by a truncated chain. The time scale, the number of oscillators in the truncated chain, and the error one makes using the effective dynamics, are strictly connected quantities.
The next Theorem proves how these quantities are related.

\vspace{0.5cm}
\noindent{\bf Theorem 3:} The error function $\epsilon(n,t)$ defined in Eqs.~\eqref{eps1}-\eqref{eps2} is bounded from above as follows:
\begin{eqnarray}\label{bounderr}
\epsilon(n,t)&\leq&\sum_{k=1}^N\left|P_{n}(\omega_k)\right|\left(\left|q_k(0)\right|+\frac{\left|\dot{q}_k(0)\right|}{\omega_k}\right)t^{2n+2}\cdot\nonumber\\
&&\left[\frac{\cosh\left(t\sqrt{\sum_{i=0}^n\Omega_i^2}\right)}{(2n+2)!}
+\frac{D^2t^4\cosh\left(t\sqrt{\Omega^2+\Omega_1^2+\sum_{i=0}^n\Omega_i^2}\right)}{(2n+6)!}\right]
\end{eqnarray}

\noindent{\bf Proof:} We apply the triangular inequality to Eq.~\eqref{err}, and we evaluate independently the two contributions of $|\epsilon_1(n,t)|$ and $|\epsilon_2(n,t)|$. The first one reads
\begin{eqnarray}
|\epsilon_1(n,t)|&\leq&\left(\prod_{l=0}^n\frac{D_l}{\Omega_l}\right)\int_0^t \left|K_{n}(t-s)\right|\cdot\left|X_{n+1}(s)\right|ds\nonumber\\
&\leq&\left(\prod_{l=0}^n\Omega_l^{-1}\right)\sum_{k=1}^N\left|P_{n}(\omega_k)\right|\left(\left|q_k(0)\right|+\frac{\left|\dot{q}_k(0)\right|}{\omega_k}\right)\int_0^t \left|K_{n}(t-s)\right|ds
\end{eqnarray}
where the second line is obtained by expressing $X_{n+1}$ in terms of the $q_k$ by means of Eqs.~\eqref{X} and \eqref{O}. Since the $q_k$ are independent, they evolve with linear combinations of sines and cosines which have been bounded by 1.
One can then exploit Eq.~\eqref{taykern} and obtain
\begin{eqnarray}
\int_0^t \left|K_{n}(t-s)\right|ds&\leq&\sum_{j=n+1}^{\infty}\frac{K_n^{(2j-1)}(0)}{2j!}t^{2j}\nonumber\\
&\leq&\left(\prod_{l=0}^i\Omega_l\right)\sum_{j=n+1}^{\infty}\frac{t^{2j}}{2j!}\left(\sum_{i=1}^n\Omega_i^2\right)^{j-n-1}\nonumber\\
&\leq& \left(\prod_{l=0}^i\Omega_l\right)\frac{t^{2n+2}}{(2n+2)!}\cosh\left(t\sqrt{\sum_{i=1}^n\Omega_i^2}\right)\,.
\end{eqnarray}
The second inequality is obtained by substituting Eq.~\eqref{k0} for $K_n^{(2j-1)}(0)$ and by observing that this equation is essentially $(\sum_{j=0}^i\Omega_j^2)^{n-1-i}$ with all the coefficients set to $1$. The third line comes from a change of variable on the sum index and extending the series to zero.
Substituting this result in Eq.~\eqref{bounderr} one finds
\begin{equation}
|\epsilon_1(n,t)|\leq\sum_{k=1}^N\left|P_{n}(\omega_k)\right|\left(\left|q_k(0)\right|+\frac{\left|\dot{q}_k(0)\right|}{\omega_k}\right)\frac{t^{2n+2}}{(2n+2)!}\cosh\left(t\sqrt{\sum_{i=1}^n\Omega_i^2}\right)\,.
\end{equation}
The contribution of $|\epsilon_2(n,t)|$ to Eq.~\eqref{bounderr} is obtained as follows. First of all one observes that the integral kernel of $\epsilon_2(n,t)$ in~\eqref{eps2} is of the same order as $K_1$:
\begin{equation}
\frac{\mu_2\sin[\mu_1(t-s)]-\mu_1\sin[\mu_2(t-s)]}{\mu_2^2-\mu_1^2}:=\tilde{K}_1(t-s)\,.
\end{equation}
Substituting the definition of $\epsilon_1(n,t)$ in Eq.~\eqref{eps2}, one finds that $X_{n+1}$ contributes via a kernel $\tilde{K}_{n+2}$ of the order $n+2$. The final result is obtained following the same procedure as for $|\epsilon_1(n,t)|$. $\square$

As expected Eq.~\eqref{bounderr} displays a dependence on the initial conditions of the bath position operators. Since these quantities are in principle not known, to overcome this issue one traces over the bath degrees of freedom. We assume that the bath initial state is in thermal equilibrium at temperature $T$:
\begin{equation}\label{therm}
\rho=\frac{1}{Z}e^{-\beta H}
\end{equation}
where $Z$ guarantees that $\mathrm{Tr}\rho=1$, and $\beta=(k_BT)^{-1}$ with $k_B$ the Boltzmann constant.
We prove the following Theorem:

\vspace{0.5cm}
\noindent{\bf Theorem 4:} If the initial state of the bath is given by Eq.~\eqref{therm}, the error is in average bounded from above as follows:
\begin{eqnarray}\label{boundav}
\langle\epsilon(n,t)\rangle&\leq&\sqrt{\frac{8k_BT}{\pi}}\left(\sum_{k=1}^N\frac{|P_{n}(\omega_k)|}{\omega_k}\right)t^{2n+2}\cdot\nonumber\\
&&\left[\frac{\cosh\left(t\sqrt{\sum_{i=0}^n\Omega_i^2}\right)}{(2n+2)!}
+\frac{D^2t^4\cosh\left(t\sqrt{\Omega^2+\Omega_1^2+\sum_{i=0}^n\Omega_i^2}\right)}{(2n+6)!}\right]
\end{eqnarray}


\noindent{\bf Proof:}  We consider the term in Eq.~\eqref{bounderr} that depends on the initial operators, and we average it over the initial state~\eqref{therm}:
\begin{equation}
\sum_{k=1}^N\frac{\left|P_{n}(\omega_k)\right|}{Z}\int_{-\infty}^{\infty}\prod_{l=1}^N dq_l d\dot{q}_l\left(\left|q_k(0)\right|+\frac{\left|\dot{q}_k(0)\right|}{\omega_k}\right)\exp\left(-\frac{\beta}{2}\sum_{j=1}^N\dot{q}_j^2+\omega_j^2q_j^2\right)\,.
\end{equation}
One can easily perform the two $N$-dimensional Gaussian integrals and obtain
\begin{equation}
\sqrt{\frac{8k_BT}{\pi}}\sum_{k=1}^N\frac{\left|P_{n}(\omega_k)\right|}{\omega_k}
\end{equation}
Replacing this expression in Eq.~\eqref{bounderr} one finally obtains Eq.~\eqref{boundav}. $\square$

Theorem 4 shows how $\epsilon$, $t$ and $n$ relate to each other at a given temperature. This is a fundamental result as it explains in which sense and under which limits the truncated dynamics can be considered an effective dynamics for the full system. Moreover, the result is versatile and can be easily adapted to different uses, since one can fix two of the mentioned quantities in order to obtain the third one. For example, one might be interested to know how big is the error after a time $t$, truncating the chain to $n$ oscillators. On the other side, one might choose to fix a value for the error at a given time, and derive how many oscillators are needed in order to satisfy such condition. It is not easy to invert analytically Eq.~\eqref{boundav}, but it is rather easy to evaluate it numerically.

\section{Conclusions}
In order to derive an effective description of non-Markovian open systems dynamics, we considered a chain representation of the environment. After deriving analytically the exact dynamics of the problem,  we proved the existence of an effective description for such a dynamics and we provided its explicit form. The proof is based on a peculiar feature of the integral kernels of the dynamics, deeply related to the harmonic feature of the model. In particular, this intrinsic structure shows that \lq\lq far\rq\rq~oscillators can be neglected if the timescale of the dynamics is short enough.
Furthermore, we gave an upper bound of the error one makes when considering such an approximated dynamics instead of the exact one. Such bound depends on the number of oscillators in the truncated chain, on the timescale and on the temperature of the system, setting a strong relationship among these quantities. This achievement gives a strong basis for future investigations on non-Markovian systems both at the analytical and numerical levels. For example, our result provides a way to determine the number of modes in a chain necessary to keep the error small, or the timescale at which a numerical simulation can be considered accurate.

\section{Acknowledgements}
The work of LF was supported by the Marie Curie Fellowship PIEF-GA-2012-328600. The authors wish to thank A. Bassi for many valuable discussions.


\begin{thebibliography}{99}

\bibitem{BrePet02} H.P. Breuer and F. Petruccione Theory of open quantum systems (Oxford, Oxford University Press, 2002).

\bibitem{Pometal} A. Pomyalov {\it et al.} J. Chem. Phys. {\bf 123}, 204111 (2005); 
                  T. Gu\'erin {\it et al.} Nat. Chem. {\bf 4}, 568 (2012).
\bibitem{Xiaetal}  X.Q. Xiang {\it et al.}, Phys. Rev. A {\bf 83}, 053823 (2011); 
                  B.D. Fainberg {\it et al.}, Phys. Rev. B {\bf 83}, 205425 (2011).
\bibitem{Thoetal} M. Thorwart {\it et al.}, Chem. Phys. Lett. {\bf 478}, 234 (2009); 
                  X.-T. Liang, Phys. Rev. E {\bf 82}, \mbox{051918 (2010);} P. Nalbach {\it et al.}, New J. Phys. 12 065043 (2010);
                  T. Scholak {\it et al.}, Europhys. Lett. 96, 10001 (2011); 
                  P. Nalbach {\it et al.}, Phys. Rev. E {\bf 84}, 041926 (2011); 
                  G.D. Scholes {\it et al.}, Nature Chem. {\bf 3}, 763 (2011).


\bibitem{Zwa73} R. Zwanzig, J. Stat. Phys. {\bf 9}, 215 (1973).

\bibitem{ForLewOco88a} G. W. Ford, J. T. Lewis, R. F. O'Connell, J. Stat. Phys. {\bf 53}, 439 (1988).

\bibitem{CalLeg83} A.O. Caldeira and A. Leggett, Physica A {\bf 121}, 587 (1983). 

\bibitem{GraTal83} H. Grabert, P. Talkner, Phys. Rev. Lett. {\bf 50} , 1335 (1983).

\bibitem{HuPazZha92} B. L. Hu, J. P. Paz and Y. Zhang, Phys. Rev. D {\bf 45}, 2843 (1992).


\bibitem{Mor65} H. Mori, Prog. Theor. Phys. {\bf 33}, 423 (1965): Prog. Theor. Phys. {\bf 34}, 399 (1965).

\bibitem{Kub66} R. Kubo, Rep. Prog. Phys. {\bf 29}, 255 (1966).

\bibitem{BenKac81} R. Benguria, M. Kac, Phys. Rev. Lett. {\bf 46}, 1 (1981). 

\bibitem{ForKac87} G. W. Ford, M. Kac, J. Stat. Phys. {\bf 46}, 803 (1987).

\bibitem{ForLewOco88b} G. W. Ford, J. T. Lewis, and R. F. O'Connell, Phys. Rev. A {\bf 37}, 4419 (1988).


\bibitem{FerDur15} L. Ferialdi, D. D\"urr, Phys. Rev. A {\bf 91}, 042130 (2015).

\bibitem{Rub60} R. J. Rubin, J. Math. Phys. {\bf 1}, 309 (1960).

\bibitem{ForKacMaz65} G. W. Ford, M. Kac, P. Mazur, J. Math. Phys. {\bf 6}, 504 (1965).

\bibitem{GinBurCed06} E. Gindensperger, I. Burghardt, L. S. Cederbaum, J. Chem. Phys. {\bf 124}, 144103 (2006).

\bibitem{Gla04} G. M. L. Gladwell, {\it Inverse Problem in Vibration}, Kluwer Academic Publishers, (2004).

\bibitem{ChuGol05}M. T. Chu, G. H. Golub, {\it Inverse Eigenvalue Problems}, Oxford Science Publications, (2005). 

\bibitem{PolMan08}
A.~D. Polyanin and A.~V. Manzhirov, \emph{Handbook of integral equations},
  Chapman \& Hall/CRC, Boca Raton, FL (2008).







\end{thebibliography}
\end{document}